\overfullrule=0pt
\input harvmac

\def\p{\partial}
\def\pb{\overline\p}

\def\a{\alpha}
\def\b{\beta}
\def\g{\gamma}
\def\d{\delta}
\def\e{\epsilon}
\def\r{\rho}

\def\th{\theta}
\def\l{\lambda}
\def\o{\omega}

\def\Mb{\overline M}
\def\Nb{\overline N}
\def\Pb{\overline P}

\def\Jb{\overline J}
\def\O{\Omega}
\def\Ab{\overline A}
\def\Xt{\widetilde X}
\def\tht{\widetilde\th}
\def\N{\nabla}
\def\NN{\overline\N}
\def\Db{\overline\Delta}
\def\vphi{\varphi}

\Title{\vbox{}} {\vbox{ \centerline{\bf A Note on T-dualities in the Pure Spinor Heterotic String}}}
\bigskip\centerline{Osvaldo Chand\'{\i}a\foot{e-mail: ochandia@unab.cl}}
\bigskip
\centerline{\it Departamento de Ciencias Fisicas }
\centerline{\it Universidad Andres Bello, Santiago, Chile}

\vskip .3in

In this note we study the preservation of the classical pure spinor BRST constraints under super T-duality transformations. We also determine the invariance of the one-loop conformal invariance and of the local gauge and Lorentz anomalies under the super T-dualities.

\vskip .3in

\Date{February 2008}

\newsec{Introduction}

Dualities in string theory have become a very important property
in string theory. They allow to show equivalence between different types of strings implying, in this way, a unifying criteria in string theory. One of these dualities is the target (T-)duality which makes type IIA equivalent to type IIB \ref\tduality{ A.~Giveon, M.~Porrati and E.~Rabinovici, ``Target Space Duality in String
Theory,'' Phys.\ Rept.\  244 (1994) 77 [arXiv:hep-th/9401139].}.
From the world-sheet point of view, T-dualities imply equivalence between different types of string theory backgrounds following the Buscher procedure \ref\BuscherQJ{ T.~H.~Buscher, ``Path Integral
Derivation of Quantum Duality in Nonlinear Sigma Models,'' Phys.\ Lett.\  B201 (1988) 466. }. The idea is to consider a background independent of some direction and then introduce a gauge field for it. By integrating out these gauge fields, we can obtain the T-dual background. All this work nicely because it is done in the bosonic string theory. In the superstring theory case, the situation is more involved because is difficult to work with Ramond backgrounds. This difficulty is avoided if we have a superstring theory formalism similar to the bosonic string. Such formalism is claimed to be the pure spinor formalism \ref\nb1{ N.~Berkovits,
``Super-Poincare Covariant Quantization of the Superstring,''
JHEP 0004 (2000) 018 [arXiv:hep-th/0001035].}.
This covariant formalism for the superstring was invented some time ago by Berkovits. It uses the superspace coordinates as basic free world-sheet variables. Since this system is not conformal invariant, it is needed to introduce new bosonic variables $\l^\a$ constrained to satisfy $(\l\g^m\l) =0$, where $\g^m$ are the symmetric $16\times 16$ gamma matrices in ten dimensions. These variable are named as pure spinors. Conformal invariance is not enough to quantize this sting theory. The new ingredient is to postulate the nilpotent charge $Q = \oint \l^\a d_\a$ (with $d_\a$ being the world-sheet generator of translations in superspace) as
BRST charge of the system. Although it is necessary to break ten-dimensional Poincare invariance to solving the pure spinor constraint, it can be shown that the physical spectrum \ref\massive{ N.~Berkovits and O.~Chandia, ``Massive Superstring
Vertex Operator in D = 10 Superspace,'' JHEP 0208 (2002) 040
[arXiv:hep-th/0204121].} and scattering amplitudes are manifestly super Poincare invariant \ref\amplitudes{ N.~Berkovits and B.~C.~Vallilo,
``Consistency of Super-Poincare Covariant Superstring Tree Amplitudes,''
JHEP 0007 (2000) 015 [arXiv:hep-th/0004171]\semi N.~Berkovits,
``Multiloop Amplitudes and Vanishing Theorems Using the Pure Spinor Formalism for the Superstring,'' JHEP 0409 (2004) 047
[arXiv:hep-th/0406055]\semi N.~Berkovits,
``Super-Poincare Covariant Two-loop Superstring Amplitudes,''
JHEP 0601 (2006) 005 [arXiv:hep-th/0503197]\semi N.~Berkovits,
C.~R.~Mafra, ``Equivalence of Two-loop Superstring Amplitudes in the Pure Spinor and  RNS Formalisms,'' Phys.\ Rev.\ Lett.\  96 (2006) 011602 [arXiv:hep-th/0509234].}.

Strings in curved backgrounds can be constructed in this formalism  \ref\BerkovitsUE{ N.~Berkovits and
P.~S.~Howe, ``Ten-dimensional Supergravity Constraints from the
Pure Spinor Formalism for the Superstring,'' Nucl.\ Phys.\  B635
(2002) 75 [arXiv:hep-th/0112160].}, where it was shown that BRST invariance implies that the background fields satisfy the corresponding ten-dimensional supergravity constraints. It was shown that this system preserves one-loop conformal invariance in the heterotic \ref\ChandiaHN{ O.~Chandia and
B.~C.~Vallilo, ``Conformal Invariance of the Pure Spinor
Superstring in a Curved Background,'' JHEP 0404 (2004) 041
[arXiv:hep-th/0401226].} and type II \ref\Chandia{ O.~A.~Bedoya and O.~Chandia, ``One-loop Conformal Invariance of the Type II Pure
Spinor Superstring in a Curved Background,'' JHEP 0701 (2007) 042
[arXiv:hep-th/0609161].} strings as consequence of the
classical BRST constraints found in \BerkovitsUE. It was also shown in \ref\mario{ O.~Chandia and M.~Tonin, ``BRST Anomaly and Superspace Constraints of the Pure Spinor Heterotic String
in a Curved Background,'' JHEP 0709 (2007) 016 [arXiv:0707.0654 [hep-th]].} that the quantum BRST invariance is modified
consistently after using cohomological methods\foot{In \ref\BedoyaMZ{ O.~A.~Bedoya, ``Yang-Mills Chern-Simons Corrections from the Pure Spinor Superstring,'' JHEP 0809 (2008) 078
[arXiv:0807.3981 [hep-th]].}, the gauge field contribution to
the one-loop BRST invariance of the effective action was obtained.}.

In this note we will study the quantum preservation of the bosonic
\ref\BenichouIT{ R.~Benichou, G.~Policastro and J.~Troost,
``T-duality in Ramond-Ramond Backgrounds,'' Phys.\ Lett.\  B661
(2008) 192 [arXiv:0801.1785 [hep-th]].} and fermionic
\ref\BerkovitsIC{ N.~Berkovits and J.~Maldacena, ``Fermionic
T-Duality, Dual Superconformal Symmetry, and the Amplitude/Wilson
Loop Connection,'' arXiv:0807.3196 [hep-th].} T-dualities in the
pure spinor formalism. Note that a combination of bosonic and
fermionic T-dualities was used in \BerkovitsIC\ to show that the AdS$_5\times$S$^5$ background of type IIB string theory remains invariant explaining in this way the so called ``dual superconformal symmetry'' of certain planar scattering amplitudes 
in $N=4, d=4$ SYM theory \ref\DrummondVQ{ J.~M.~Drummond, J.~Henn, G.~P.~Korchemsky and E.~Sokatchev, ``Dual Superconformal Symmetry 
of Scattering Amplitudes in N=4 Super-Yang-Mills Theory,'' 
arXiv:0807.1095 [hep-th].} \ref\RicciEQ{ R.~Ricci, A.~A.~Tseytlin
and M.~Wolf, ``On T-Duality and Integrability for Strings on AdS Backgrounds,'' JHEP 0712 (2007) 082 [arXiv:0711.0707 [hep-th]]
\semi N.~Beisert, R.~Ricci, A.~A.~Tseytlin, M.~Wolf, 
``Dual Superconformal Symmetry from AdS$_5\times$S$^5$ Superstring Integrability,'' Phys.\ Rev.\  D78 (2008) 126004 
[arXiv:0807.3228 [hep-th]].} (see also the review 
\ref\sym{ L.F. Alday and R. Roiban, ``Scattering Amplitudes, Wilson Loops and the String/Gauge Theory Correspondence,'' arXiv:0807.1889 [hep-th].}).

In the next section we will review the bosonic T-duality of
\BenichouIT\ and the fermionic duality of \BerkovitsIC\ for the
heterotic superstring. Then we will explicitly check that the
classical BRST constraints are preserved after a T-duality transformation. Finally we will prove that quantum conformal and
quantum local symmetries are preserved under T-duality.

\newsec{T-duality for the Heterotic Pure Spinor Superstring}

We review the T-dualities discovered in \BenichouIT\ and
\BerkovitsIC\ for the heterotic string case. The sigma model action in this case is given by

\eqn\action{ S = {1\over{2\pi\a'}}\int d^2z ~ \ha \p Z^{\Mb} \pb
Z^{\Nb} ( G_{\Nb\Mb}(Z) + B_{\Nb\Mb}(Z) ) + \p Z^{\Mb} \Jb^I
A_{I\Mb}(Z) + d_\a \pb Z^{\Mb} E_{\Mb}{}^\a(Z) }
$$
+ d_\a \Jb^I W_I^\a(Z) + \l^\a \o_\b \pb Z^{\Mb} \O_{\Mb\a}{}^\b(Z)
+ \l^\a \o_\b \Jb^I U_{I\a}{}^\b(Z) + {\cal L}(\Jb, \l, \o) + {\cal
L}_{FT} ,$$ where $Z^{\Mb}$ are the coordinates of the
ten-dimensional heterotic superspace, $\Jb^I$ is the current for the
heterotic fermions, $d_\a$ is the world-sheet generator for
superspace translations, $\l^\a$ is the pure spinor and $\o_\a$ is
its conjugate momentum. The term ${\cal L}(\Jb, \l, \o)$ is the
action for the pure spinor variables and heterotic fermions in flat space. The term ${\cal L}_{FT}$ is the Fradkin-Tseytlin term and it is given by

\eqn\ft{ {\cal L}_{FT} = \a' {\cal R} \Phi(Z) ,} where ${\cal R}$ is
the two-dimensional scalar curvature and $\Phi$ is the dilaton
superfield. The background fields $G, B, ...$ satisfy the
supergravity equations of motion as consequence of the BRST
invariance of \action. This symmetry is generated by the nilpotent
charge $Q = \oint \l^\a d_\a$.

Now we will perform a combination of bosonic T-duality, as in
\BenichouIT, and then a fermionic T-duality, as in \BerkovitsIC. In
both cases, we will see that the action \action\ preserves its form.

\subsec{ Bosonic T-duality in the pure spinor string }

We assume that the background fields are independent of some bosonic
direction, say $X^1$, and we split $Z^{\Mb} = ( X^1, Y^M )$. As it
was noted in \BuscherQJ, the T-dual action is obtained after
gauging the $X^1$-direction by introducing a purely gauge fields $A$ and $\Ab$ as

\eqn\actiona{ S = {1\over{2\pi\a'}}\int d^2z ~ \ha A\Ab G_{11}(Y) +
\ha A \pb Y^M L_{M1}(Y) + \ha \p Y^M \Ab L_{1M}(Y) + \ha \p Y^M \pb
Y^N L_{NM}(Y) }
$$
+ A \Jb^I A_{I1}(Y) + \p Y^M \Jb^I A_{IM}(Y) + d_\a \Ab E_1{}^\a(Y)
+ d_\a \pb Y^M E_M{}^\a(Y) + d_\a \Jb^I W_I^\a(Y) $$
$$
+ \l^\a \o_\b \Ab \O_{1\a}{}^\b(Y) + \l^\a \o_\b \pb Y^M
\O_{M\a}{}^\b(Y) + \l^\a \o_\b \Jb^I U_{I\a}{}^\b(Y) + \ha \Xt^1 (
\p \Ab - \pb A ) + {\cal L}(\Jb, \l, \o) + {\cal L}_{FT} ,$$ where
$L_{\Nb\Mb} = G_{\Nb\Mb} + B_{\Nb\Mb}$ and $\Xt^1$ is a lagrange
multiplier which enforces the pure gauge condition. In fact, if we
integrate out this field, we recover the original action \action\ if

$$
A = \p X^1,\quad \Ab = \pb X^1 .$$ In order to obtain the T-dual
sigma model action, we integrate out the gauge fields instead. By
varying respect to $A$ and $\Ab$, we obtain

\eqn\aabar{ A = ( \p \Xt^1 - \p Y^M L_{1M} - 2 d_\a E_1{}^\a - 2
\l^\a \o_\b \O_{1\a}{}^\b ) {1\over{G_{11}}} ,}
$$
\Ab = ( -\pb \Xt^1 - \pb Y^M L_{M1} - 2 \Jb^I A_{I1} )
{1\over{G_{11}}} .$$ We plug this in the action \actiona\ to get

\eqn\actiont{ S = {1\over{2\pi\a'}}\int d^2z ~ \ha \p \Xt^1 \pb
\Xt^1 G'_{11}(Y) + \ha \p\Xt^1 \pb Y^M L'_{M1}(Y) + \ha \p Y^M
\pb\Xt^1 L'_{1M}(Y) }
$$
+ \ha \p Y^M \pb Y^N L'_{NM}(Y) + \p\Xt^1 \Jb^I A'_{I1}(Y) + \p Y^M
\Jb^I A'_{IM}(Y) + d_\a \pb \Xt^1 E'_1{}^\a(Y) + d_\a \pb Y^M
E'_M{}^\a(Y) $$
$$
+ d_\a \Jb^I {W'}_I^\a(Y) + \l^\a \o_\b \pb\Xt^1 \O'_{1\a}{}^\b(Y) +
 \l^\a \o_\b \pb Y^M \O'_{M\a}{}^\b(Y) $$
$$
+ \l^\a \o_\b \Jb^I U'_{I\a}{}^\b(Y) + {\cal L}(\Jb, \l, \o) + {\cal
L'}_{FT} ,$$ where the transformed fields are given by

\eqn\trfields{ G'_{11} = {1\over G_{11}},\quad L'_{M1} =
{L_{M1}\over G_{11}},\quad  L'_{1M} = - {L_{1M}\over G_{11}},\quad
L'_{NM} = L_{NM} - {L_{N1}L_{1M}\over G_{11}} ,}

$$
A'_{I1} = {A_{I1}\over G_{11}},\quad A'_{IM} = A_{IM} -
{A_{I1}L_{1M}\over G_{11}},\quad E'_1{}^\a = -{E_1{}^\a\over
G_{11}},\quad E'_M{}^\a = E_M{}^\a - {L_{M1}E_1{}^\a\over G_{11}}
,$$
$$
\O'_{1\a}{}^\b = -{\O_{1\a}{}^\b\over G_{11}},\quad \O'_{M\a}{}^\b =
\O_{M\a}{}^\b - {L_{M1}\O_{1\a}{}^\b\over G_{11}} ,$$
$$
{W'}_I^\a = W_I^\a - 2{A_{I1}E_1{}^\a\over G_{11}},\quad
U'_{I\a}{}^\b = U_{I\a}{}^\b - 2 {A_{I1}\O_{1\a}{}^\b\over G_{11}}
.$$ It remains to determine the bosonic component of the
supervielbein $E_{\Mb}{}^a$. It can be obtained through the
definition

\eqn\gee{ G_{\Nb\Mb} = E_{\Nb}{}^a E_{\Mb}{}^b \eta_{ab} .}
Following \BenichouIT, we determine this supervielbein as

\eqn\eqea{ E'_{\Mb}{}^a = Q_{\Mb}{}^{\Nb} E_{\Nb}{}^a ,} where the
matrix $Q$ is determined by

\eqn\eqealfa{ E'_{\Mb}{}^\a = Q_{\Mb}{}^{\Nb} E_{\Nb}{}^\a .}
According to \trfields, the matrix $Q$ has the entries

\eqn\entriesq{ Q_1{}^1 = - {1\over G_{11}},\quad Q_1{}^M = 0,\quad
Q_M{}^1 = -{L_{M1}\over G_{11}},\quad Q_N{}^M = \d_M{}^N .} In this
way we obtain

\eqn\eprimea{ E'_1{}^a = -{E_1{}^a\over G_{11}},\quad E'_M{}^a =
E_M{}^a - {L_{M1}E_1{}^\a\over G_{11}} .} As verification, we need to
obtain the transformations given in \trfields\ for the supermetric
by using the definition of \gee\ and \eprimea. In fact, doing this

$$
G'_{11} = {1\over G_{11}},\quad G'_{1M} = {B_{M1}\over G_{11}},
\quad G'_{NM} = G_{NM} + {B_{N1}B_{M1}-G_{N1}G_{M1}\over G_{11}}
.$$ which are compatible with \trfields. Note that we need to
perform a shift in de dilaton superfield too because of the term
involving $A\Ab$ in \actiona\ is not one. As it was shown in
\BuscherQJ, the dilaton is transformed according to

\eqn\tdilaton{ \Phi' = \Phi + \log {2\pi\a'\over G_{11}} .}

We will also need the inverse of the super vielbein.
It was found that $E'_{\Mb}{}^A = Q_{\Mb}{}^{\Nb} E_{\Nb}{}^A$
where the matrix $Q$ is given in
\entriesq. The inverse $E'_A{}^{\Mb} = E_A{}^{\Nb}
(Q^{-1})_{\Nb}{}^{\Mb}$, where the entries for the matrix $Q^{-1}$
are

\eqn\qmenos{ (Q^{-1})_1{}^1 = -G_{11},\quad (Q^{-1})_1{}^M = 0,\quad (Q^{-1})_M{}^1 =
-L_{M1},\quad (Q^{-1})_M{}^N = \d_M{}^N .}

\subsec{ Fermionic T-duality }

Let us consider now a fermionic T-duality \BerkovitsIC. We assume
that the background in the sigma model action \action\ is
independent of a fermionic direction, say $\th^1$. As in the bosonic
case, we gauge this isometry by introducing a pair of fermionic
gauge field $(A, \Ab)$ and by adding a fermionic Lagrange multiplier
which enforces a pure gauge condition on the gauge fields. The
action now is

\eqn\actionf{ S = {1\over{2\pi\a'}}\int d^2z ~ \ha A\Ab B_{11}(Y) +
\ha A \pb Y^M L_{M1}(Y) + \ha \p Y^M \Ab L_{1M}(Y) + \ha \p Y^M \pb
Y^N L_{NM}(Y) }
$$
+ A \Jb^I A_{I1}(Y) + \p Y^M \Jb^I A_{IM}(Y) + d_\a \Ab E_1{}^\a(Y)
+ d_\a \pb Y^M E_M{}^\a(Y) + d_\a \Jb^I W_I^\a(Y) $$
$$
+ \l^\a \o_\b \Ab \O_{1\a}{}^\b(Y) + \l^\a \o_\b \pb Y^M
\O_{M\a}{}^\b(Y) + \l^\a \o_\b \Jb^I U_{I\a}{}^\b(Y) + \ha \tht^1 (
\p \Ab - \pb A ) + {\cal L}(\Jb, \l, \o) + {\cal L}_{FT} .$$
Integrating out the fermionic Lagrange multiplier $\tht^1$, we
obtain the action \action\ because

$$
A = \p \th^1,\quad \Ab = \pb\th^1 .$$ We now integrate the gauge
fields. The equations of motion for them determine

\eqn\aabarf{ A = ( \p \tht^1 - (-1)^M \p Y^M L_{1M} - 2 d_\a
E_1{}^\a + 2 \l^\a \o_\b \O_{1\a}{}^\b ) {1\over{B_{11}}} ,}
$$
\Ab = ( \pb \tht^1 - \pb Y^M L_{M1} - 2 \Jb^I A_{I1} )
{1\over{B_{11}}} ,$$ where $(-1)^M$ is $-1$ if $M$ is a bosonic
index and is $+1$ if $M$ is a fermionic index. We plug these values
in the action \actionf\ to obtain the fermionic T-dual background

\eqn\actiontf{ S = {1\over{2\pi\a'}}\int d^2z ~ \ha \p \tht^1 \pb
\tht^1 B'_{11}(Y) + \ha \p\tht^1 \pb Y^M L'_{M1}(Y) + \ha \p Y^M
\pb\tht^1 L'_{1M}(Y) }
$$
+ \ha \p Y^M \pb Y^N L'_{NM}(Y) + \p\tht^1 \Jb^I A'_{I1}(Y) + \p Y^M
\Jb^I A'_{IM}(Y) + d_\a \pb \tht^1 E'_1{}^\a(Y) + d_\a \pb Y^M
E'_M{}^\a(Y) $$
$$
+ d_\a \Jb^I {W'}_I^\a(Y) + \l^\a \o_\b \pb\tht^1 \O'_{1\a}{}^\b(Y)
+ \l^\a \o_\b \pb \tht^1 \O'_{1\a}{}^\b(Y) + \l^\a \o_\b \pb Y^M
\O'_{M\a}{}^\b(Y) $$
$$
+ \l^\a \o_\b \Jb^I U'_{I\a}{}^\b(Y) + {\cal L}(\Jb, \l, \o) + {\cal
L'}_{FT} ,$$ where the fermionic T-dual background is given by

\eqn\trfieldsf{ B'_{11} = -{1\over B_{11}},\quad L'_{M1} =
{L_{M1}\over B_{11}},\quad  L'_{1M} = {L_{1M}\over B_{11}},\quad
L'_{NM} = L_{NM} - {L_{N1}L_{1M}\over B_{11}} ,}
$$
A'_{I1} = {A_{I1}\over B_{11}},\quad A'_{IM} = A_{IM} -
{A_{I1}L_{1M}\over B_{11}},\quad E'_1{}^\a = {E_1{}^\a\over
B_{11}},\quad E'_M{}^\a = E_M{}^\a - {L_{M1}E_1{}^\a\over B_{11}}
,$$
$$
\O'_{1\a}{}^\b = {\O_{1\a}{}^\b\over B_{11}},\quad \O'_{M\a}{}^\b =
\O_{M\a}{}^\b - {L_{M1}\O_{1\a}{}^\b\over B_{11}} ,$$
$$
{W'}_I^\a = W_I^\a - 2{A_{I1}E_1{}^\a\over B_{11}},\quad
U'_{I\a\b} = U_{I\a}{}^\b - 2 {A_{I1}\O_{1\a}{}^\b\over B_{11}}
.$$ We note that these transformations are very similar to those
in the bosonic case. In the fermionic case, the $B_{11}$ plays the
role of $G_{11}$ in the bosonic case. It remains to determine the bosonic
component of the supervielbein. As in the bosonic case, we note
that is given by \eqea\ where the matrix $Q$ is determined by
\eqealfa\ and the transformations of \trfieldsf. The entries of
$Q$ are

\eqn\entriesqf{ Q_1{}^1 = {1\over B_{11}},\quad Q_1{}^M = 0,\quad
Q_M{}^1 = -{L_{M1}\over B_{11}},\quad Q_N{}^M = \d_M{}^N .}
Therefore,

\eqn\eprimeaf{ E'_1{}^a = {E_1{}^a\over B_{11}},\quad E'_M{}^a =
E_M{}^a - {L_{M1}E_1{}^\a\over B_{11}} .} From the definition \gee,
we determine that the transformations of the supersymmetric components are

$$
G'_{M1} = {G_{M1}\over B_{11}}, \quad G'_{NM} = G_{NM} -
{G_{N1}B_{1M}+B_{N1}G_{1M}\over B_{11}} .$$ which are compatible
with \trfieldsf. As before, the dilaton superfield is shifted as

\eqn\tdilatonf{ \Phi' = \Phi - \log {2\pi\a'\over B_{11}} .} Here
the relative sign is reverted respect to \tdilaton\ because the
measure for the gauge field is grassmannian, then the jacobian of
the transformation is inverted \BerkovitsIC.

\newsec{ Classical BRST Invariance }

In the previous section it was shown how the T-dualities are realized in the pure spinor heterotic string. The classical BRST invariance of the sigma model action puts them
background field on-shell \BerkovitsUE. Since the BRST charge $Q =
\oint \l^\a d_\a$ is unaffected by the T-duality transformations,
then the T-dual background is also a solution of the $N = 1$
ten-dimensional supergravity/SYM equations of motion. Therefore, we expect that the T-dual sigma model action, both bosonic and fermionic, is
also BRST invariant. Now it will be shown that this is the case.

The classical BRST invariance of the action implies that the
background fields satisfy the constraints \BerkovitsUE\ (see also
\ref\ChandiaIX{ O.~Chandia, ``A Note on the Classical BRST
Symmetry of the Pure Spinor String in a Curved Background,''
JHEP 0607 (2006) 019 [arXiv:hep-th/0604115].}\foot{The constraints coming from the holomorphicity of the BRST current are implied by these and Bianchi identities. }

\eqn\qtwo{ \l^\a \l^\b T_{\a\b}{}^A = \l^\a \l^\b H_{\a\b
A} = \l^\a \l^\b F_{I\a\b} = \l^\a \l^\b \l^\g R_{\a\b\g}{}^\d = 0
.}
Now it will be shown that these constraints remain after a T-dual transformation of the background. Consider the bosonic T-duality of \trfields, the fermionic case is analogous.

Let's start with $\l^\a \l^\b F_{I\a\b} = 0$. We use that

$$
F'_{I\a\b} = (-1)^{\Mb+1} E'_{\b}{}^{\Mb} E'_{\a}{}^{\Nb} F'_{I\Nb\Mb} .$$
Now we split the index $\Mb$ into $(1, M)$ and we use the definition of $F$ in terms of the gauge potential $A$ together with the transformations of \trfields\ to obtain

$$
F'_{I\a\b} = F_{I\a\b} - {1\over G_{11}} ( H_{\a\b 1} + T_{\a\b}{}^a E_1{}^b \eta_{ab} ) A_{I1} .$$
If we hit this expression with $\l^\a\l^\b$ we obtain

$$
\l^\a \l^\b F'_{I\a\b} = \l^\a \l^\b F_{I\a\b} = 0.$$

Consider now the constraint involving $H$. Starting from

$$
H'_{\a\b A} = (-1)^{\Mb + 1} E'_A{}^{\Pb} E'_\b{}^{\Nb} E'_\a{}^{\Mb} H'_{\Mb\Nb\Pb} ,$$
by splitting the index $\Mb$ as $(1, M)$, the definition of $H$ in terms of $B$ and the T-dual transformations \trfields\ we obtain

$$
H'_{\a\b A} = H_{\a\b A} + {1\over G_{11}} E_\a{}^{\Mb} G_{\Mb 1} E_1{}^a T_{\b A}{}^b \eta_{ab} = H_{\a\b A} .$$
Then, the constraint $\l^\a \l^\b H_{\a\b A} = 0$ is preserved.

Consider the torsion constraint in \qtwo. Starting from

$$
T'_{\a\b}{}^A = (-1)^{\Mb + 1} E'_\b{}^{\Mb} E'_\a{}^{\Nb} T'_{\Nb\Mb}{}^A ,$$
using the definition of the torsion as the covariant derivative of the vielbein and splitting index $\Mb$ into $(1, M)$ we obtain

$$
T'_{\a\b}{}^A = T_{\a\b}{}^A -{1\over G_{11}} ( H_{\a\b 1} + T_{\a\b}{}^a E_1{}^b\eta_{ab} ) E_1{}^A ,$$
from which we obtain

$$
\l^\a \l^\b T'_{\a\b}{}^A = \l^\a \l^\b T_{\a\b}{}^A .$$

Finally, consider the torsion constraint in \qtwo. Starting from

$$
R'_{\a\b\g}{}^\d = (-1)^{\Mb+1} E'_\b{}^{\Mb} E'_\a{}^{\Nb} R'_{\Nb\Mb\g}{}^\d ,$$
using the definition of $R$ in terms of the connection $\O$ and splitting the index $\Mb$ into $(1, M)$ we obtain

$$
R'_{\a\b\g}{}^\d = R_{\a\b\g}{}^\d - {\O_{1\g}{}^\d \over G_{11}}( H_{\a\b 1} + T_{\a\b}{}^a E_1{}^b \eta_{ab} ) ,$$
then, we get

$$
\l^\a \l^\b R'_{\a\b\g}{}^\d = \l^\a \l^\b R_{\a\b\g}{}^\d .$$

\newsec{ Quantum super T-duality }

In this section we discuss the preservation of some symmetries of the sigma model action \action\ under background T-duality transformations. We consider the conformal, gauge and local Lorentz symmetries

\subsec{ One-loop conformal invariance }

As we have already stated, the classical BRST invariance of the sigma model action puts the background fields on-shell \BerkovitsUE\. Since the BRST charge $Q =
\oint \l^\a d_\a$ is unaffected by the T-duality transformations,
then the T-dual background is also a solution of the $N = 1$
ten-dimensional supergravity/SYM equations of motion. As it was
shown in \ChandiaHN, the
one-loop conformal invariance of the sigma model action is a
consequence of the classical BRST invariance. Therefore, we expect
that the T-dual sigma model action, both bosonic and fermionic, is
also BRST invariant. Now it will be shown that this is the case.

The classical BRST invariance of the action implies that the
background fields satisfy the constraints \qtwo. Recall that the $\O$ connection here involves the usual Lorentz
connection and a connection for the scaling invariance of \action\
through

\eqn\conn{ \O_{M\a}{}^\b = \O_M \d_\a{}^\b + {1\over 4}
(\g^{ab})_\a{}^\b \O_{Mab} .}

After performing a covariant background field expansion, it was
shown in \ChandiaHN\ that the one-loop UV divergence of the
effective action vanishes after using the constraints of
\qtwo, Bianchi identities and the relation

\eqn\dilom{ \N_\a \Phi = 4~\O_\a ,} where $\Phi$ is a superfield
which appears in the Fradkin-Tseytlin term in \ft\ and $\O_\a =
E_\a{}^M \O_M$. The relation \dilom\ can be obtained by requiring
the vanishing of the ghost number anomaly as it was discussed in
\BerkovitsUE\ and \Chandia.

In the T-dual background, the BRST constraints are equivalent to
\qtwo\ but with the curvatures constructed the T-dual
connections of \trfields\ in the bosonic case or the connections
of \trfieldsf\ in the fermionic case. Now one can perform a
covariant background field expansion and demonstrate the vanishing
of the one-loop effective action. In order to do this, it is
necessary to use the relation \conn\ in the T-dual background,
that is

\eqn\dilomf{ \N'_\a \Phi' = 4~\O'_\a ,} where the covariant
derivative is defined with the T-dual transformed connections of
\trfields\ or \trfieldsf, and $\Phi'$ is given by \tdilaton\ or
\tdilatonf. There is an apparent contradiction because if one uses
the rules for transforming the background field under a T-dual
transformation, the rhs of \dilomf\ remains invariant. Let us
prove this assertion in the bosonic case. In the fermionic case,
the proof can be done in parallel.

Using this,

$$
\O'_\a = E'_\a{}^{\Mb} \O'_{\Mb} = - G_{11} E_\a{}^1 \O'_1 -
E_\a{}^N L_{N1} \O'_1 + E_\a{}^M \O'_M = \O_\a.$$ Analogously, the
lhs of \dilomf\ transforms as

$$
\N'_\a \Phi' = \N_\a \Phi - {1\over G_{11}} \N_\a G_{11} .$$ By
combining both transformations, we find a contradiction with
\dilom. This is solved by recalling that the action of \action\ is
invariant under a scaling transformations where $\d\l^\a =
\e\l^\a, \d\o_\a = -\e\o_\a,\dots$ and $\d \O_M = -\p_M \e$. Using
this symmetry, we just change $\O_M \to \O_M - {1\over 4} \N_M
\log G_{11}$ to preserve the equation \dilom.

\subsec{ Gauge and local Lorentz symmetries }

The action of the pure spinor heterotic string \action\ is invariant under local gauge and local Lorentz transformations. Under the former, the background field $A_{I\Mb}$ transforms as a connection and all other fields carrying an index $I$ transform in the adjoint representation of the gauge group. Of course, fields without an index $I$ are inert under the gauge group. Under a local Lorentz transformation, $\O$ is the connection and all other fields transform homogenously.

It is well known that the effective action is potentially anomalous because these symmetries act on chiral fermions. Consider first the heterotic fermions in the action of \action. The effective action is determined by performing a covariant background expansion and then integrating out the quantum fluctuations (see \ref\skenderis{ J.~de Boer and K.~Skenderis, ``Covariant Computation of the Low
Energy Effective Action of the Heterotic Superstring,''
Nucl.\ Phys.\  B481 (1996) 129 [arXiv:hep-th/9608078].} and
references therein). It turns out the part of the effective action involving the heterotic fermions is given by

\eqn\seffa{ e^{-S_{eff}[a]} = \int D\r ~ e^{ {1\over{4\pi\a'}} \int d^2z ~ Tr (\r\N\r) } ,}
where the trace is over the vector representation of the gauge group, the heterotic fermions belong to this representation, the covariant derivative is given by $\N\r = \p\r + a\r$, where $a = a_I K^I$ with $K^I$ denoting the generators of the gauge group and

\eqn\aaa{ a_I = \p Z^{\Mb} A_{I\Mb} + d_\a W^\a_I + \l^\a \o_\b U_{I\a}{}^\b ,}
which is determined in the covariant background field expansion performed in \ChandiaHN. The gauge anomaly is given by making a gauge transformation with the gauge field in the combination of \aaa. Now it will be shown that this world-sheet field is invariant under the bosonic T-duality \trfields. In fact,

$$
a'_I = \p\Xt^1 A'_{I1} + \p Y^M A'_{IM} + d_\a W'^\a_I + \l^\a \o_\b U'_{I\a}{}^\b = \p\Xt^1 {A_{I1}\over G_{11}} + \p Y^M ( A_{IM} - {A_{I1}L_{1M}\over G_{11}} )$$
$$
+ d_\a ( W^\a_I - 2 {A_{I1}E_1{}{}^\a \over G_{11} } ) + \l^\a \o_\b ( U_{I\a}{}^\b - 2 {A_{I1}\O_{1\a}{}^\b \over G_{11} } ) ,$$
by using the equations \aabar\ and that $A = \p X^1, \Ab = \pb X^1$, we obtain that $a'_I = a_I$. Therefore, the anomalies determined by this world-sheet field are invariant under the bosonic T-duality of \trfields.

Consider now the fermionic T-duality case. We obtain

$$
a'_I = \p\tht^1 A'_{I1} + \p Y^M A'_{IM} + d_\a W'^\a_I + \l^\a \o_\b U'_{I\a}{}^\b = \p\tht^1 {A_{I1}\over B_{11}} + \p Y^M ( A_{IM} - {A_{I1}L_{1M}\over B_{11}} )$$
$$
+ d_\a ( W^\a_I - 2 {A_{I1}E_1{}\a \over B_{11} } ) + \l^\a \o_\b ( U_{I\a}{}^\b - 2 {A_{I1}\O_{1\a}{}^\b \over B_{11} } ) ,$$
by using the equations \aabarf\ and that $A = \p \th^1, \Ab = \pb \th^1$, we obtain that $a'_I = a_I$. Therefore, we have verified that the T-dualities do not affect the gauge anomaly in the heterotic string case

Now we consider the local Lorentz symmetry of the action \action. We have two potentially anomalous chiral systems, one coming from the expansion of $(d_\a, Z^{\Mb} E_{\Mb}{}^\b)$ and the other from the expansion of the pure spinor ghosts $(\o_\a, \l^\b)$. In both cases, the effective action is of the form

\eqn\sefflorentz{ e^{-S_{eff}[\Db]} = \int D\psi D\vphi ~ e^{-{1\over{2\pi\a'}} \int d^2z ~ \psi_\a \NN \vphi^\a } ,}
where $(\vphi_\a , \psi_\b)$ is $(Z^{\Mb} E_{\Mb}{}^\a , d_\b)$ or $(\l^\a , \o_\b)$ and the covariant derivative is given by $\NN\vphi^\a = \pb\vphi^\a + \vphi^\b \Db_\b{}^\a$ with the connection given by

\eqn\conneff{ \Db_\a{}^\b = \pb Z^{\Mb} \O_{\Mb\a}{}^\b + \Jb^I U_{I\a}{}^\b .}

Now it will be shown that \conneff\ is invariant under T-dualities. Using the rules of \trfields\ for the bosonic T-duality we obtain

$$
\Db'_\a{}^\b = \pb \Xt^1 \O'_{1\a}{}^\b + \pb Y^M \O'_{M\a}{}^\b + \Jb^I U'_{I\a}{}^\b  = - \pb \Xt^1 {\O_{1\a}{}^\b \over G_{11} } + \pb Y^M ( \O_{M\a}{}^\b - {L_{M1}\O_{1\a}{}^\b \over G_{11}} )$$
$$
+ \Jb^I ( U_{I\a}{}^\b - 2 {A_{I1}\O_{1\a}{}^\b \over G_{11}} ) = \Db_\a{}^\b ,$$
where we have used the equations of \aabar\ and $\Ab = \pb X^1$.

Analogously, the fermionic T-duality also leaves invariant the world-sheet field $\Db$. Then, we can conclude that the anomalies which depend on $\Db$, as the local Lorentz anomaly, will not be affected by T-dualities.

\vskip 20pt {\bf Acknowledgements:} I would like to thank Nathan Berkovits for useful comments and suggestions. This work is supported by FONDECYT grant 1061050 and Proyecto Interno DI-03-08/R from UNAB.

\listrefs

\end